\documentclass[12pt,a4paper]{amsart}
\usepackage{amssymb,amsfonts,amsthm,amsmath,graphicx,xcolor,csquotes}
\usepackage[scr]{rsfso}
\usepackage{chngcntr}
\usepackage{enumitem}

\theoremstyle{plain}
\newtheorem{theorem}{Theorem}

\newtheorem{definition}{Definition}

\newtheorem{corollary}{Corollary}

\numberwithin{theorem}{section}
\numberwithin{equation}{section}
\numberwithin{statement}{section}
\numberwithin{lemma}{section}
\numberwithin{definition}{section}
\numberwithin{conjecture}{section}
\numberwithin{corollary}{section}
\begin{document}
\title[Rogers-Ramanujan identities in Statistical Mechanics]{Rogers-Ramanujan identities in Statistical Mechanics}
\author{Geoffrey B Campbell}
\address{Mathematical Sciences Institute,
         The Australian National University,
         Canberra, ACT, 0200, Australia}

\email{Geoffrey.Campbell@anu.edu.au}



\begin{abstract}
We describe the story of the Rogers-Ramanujan identities; being known for 85 years and having about 130 pure mathematics proofs, suddenly entering physics when Rodney Baxter solved the Hard Hexagon Model in Statistical Mechanics in 1980. We next cover the accompanying proofs by George E Andrews of other related Baxter identities arisen of Rogers-Ramanujan type, leading into a new flourishing partnership of Physics and Mathematics. Our narrative goes into the subsequent 44 years, explaining the progress in physics and mathematical analysis. Finally we show some related crossovers with regard to the \textit{Elliptic q-gamma function} and some \textit{Vector Partition} generating functional equations; the latter of which may be new. The present paper is essentially chapter 11 of the author's 32 chapter book \cite{gC2024} to appear in June 2024.
\end{abstract}

\maketitle

\section{Baxter's Hard Hexagon Model solved exactly by Rogers-Ramanujan identities}\index{Rogers-Ramanujan identities}\index{Ramanujan, S.}

The years 1980 and 1981 were amazing in the theory of partitions; with significance arisen from a remarkable discovery in theoretical physics. In 1980 Baxter \cite{rB1980} found real-world physical application of the Rogers-Ramanujan identities\index{Rogers-Ramanujan identities}\index{Ramanujan, S.} coming directly from his solving the Hard Hexagon Model\index{Hard hexagon model} in Statistical Mechanics. This opened up a new world of research in both theoretical physics and partition-theoretic mathematics that is still evolving, over forty years later. This discovery has given rise to thousands of mathematics and physics research papers and promises continued relevance into future decades.


Some of this excitement can be gleaned from the important paper by Andrews\index{Andrews, G.E.} in 1981 (see \cite{gA1981}) from quoting his opening paragraph here.

\bigskip

\textit{Think About It...}

\begin{quote}
“In 1980, Baxter found his beautiful solution to the hard-hexagon
model\index{Hard hexagon model} of statistical mechanics. His treatment of this model
is naturally divided into six regimes that depend on values
taken by various parameters associated with the model. Then
in truly astounding fashion it turns out that eight
Rogers-Ramanujan type identities\index{Rogers-Ramanujan identities}\index{Ramanujan, S.}, all essentially known to
Rogers (\cite{lR1894}, \cite{lR1917}), are the fundamental keys for finding infinite product
representations of the related statistical mechanics partition
functions in \textbf{regimes I, III, and IV}.''

\bigskip

\textit{- George E. Andrews, Proceedings of the National Academy of Sciences, 1981}\index{Quotes:Think about it!Andrews, G.E.}\index{Baxter, R.J.}\index{Andrews, G.E.}\index{Baxter, R.J.}
\end{quote}

\subsection{A few words about the types of lattice models studied here.}

\begin{figure} [ht]
\centering
    \includegraphics[width=14.5cm,angle=0,height=6.37cm]{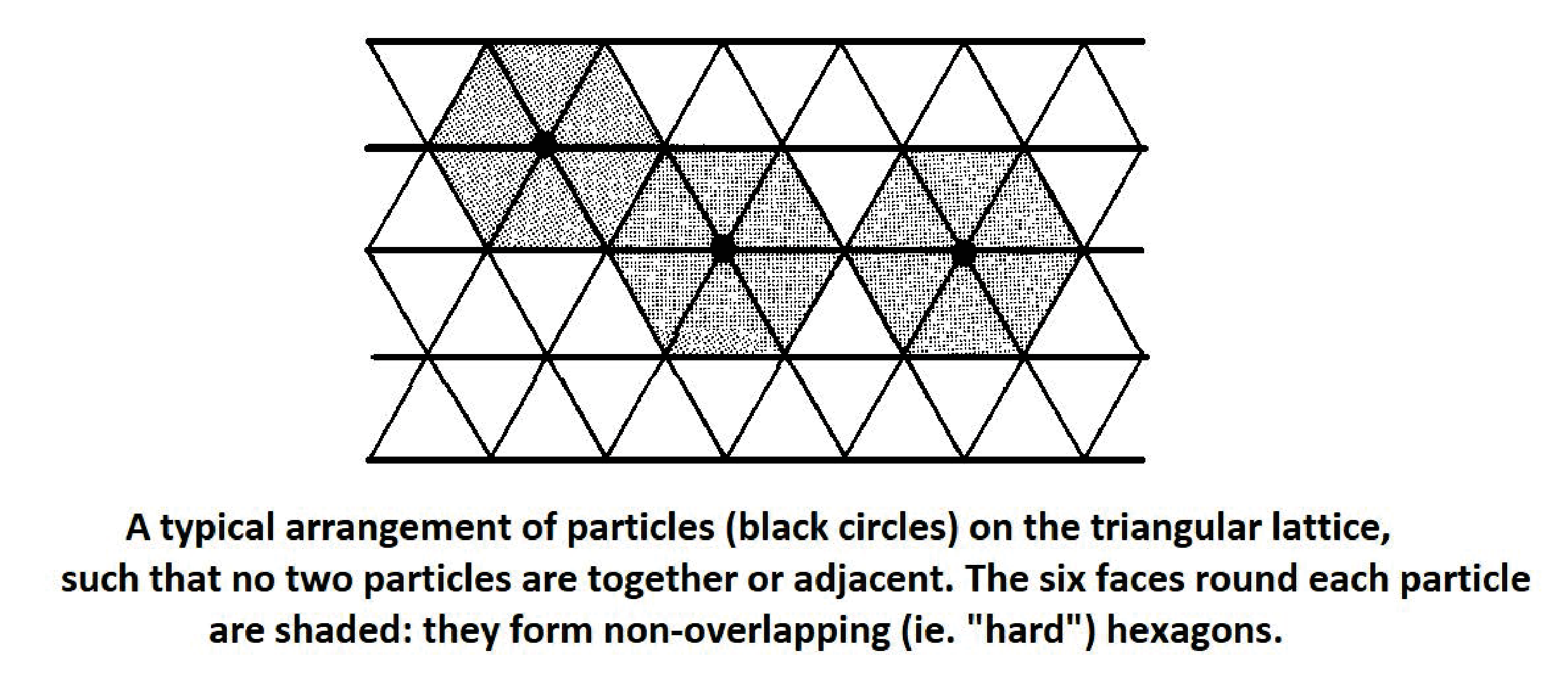}
  \caption[The particle arrangement for the Hard Hexagon Model.]{The particle arrangement for the Hard Hexagon Model.} \label{Fig6}
\end{figure}

Baxter formulates his approach based on the \textit{hard hexagon} diagram (see next page Figure \ref{Fig6}) which has constraints on the particles such that they can only be in certain places at certain moments.

The hard hexagon model\index{Hard hexagon model} is a two-dimensional lattice model of a gas of hard (i.e. non-overlapping) molecules.
In it, particles are placed on the sites of the triangular lattice so that no
two particles are together or adjacent. A typical allowed arrangement of
particles is shown in Figure \ref{Fig6}. If we regard each particle as the centre of
a hexagon covering the six adjacent faces (such hexagons are shown shaded
in the figure), then the rule only allows hexagons that do not overlap: hence
the name of the model.
For a lattice of $N$ sites, the grand-partition function is

\begin{equation} \label{7.01}
  Z(z) = \sum_{n=0}^{n/3} z^n g(n,N)
\end{equation}
where $g(n, N)$ is the allowed number of ways of placing $n$ particles on the
lattice, and the sum is over all possible values of $n$. (Since no more than
$\frac{1}{3}$ of the sites can be occupied, $n$ takes values from $0$ to $\frac{N}{3}$.)

 \index{Hard hexagon model}

Clearly, from the mathematical `non-physicist' perspective, that this schematic diagram could lead to enumeration of well defined types of integer partitions was (back in 1980 at least) seen as a major departure from the more typical Ferrers Graphs, Durfee Squares\index{Durfee square} and Plane Partition\index{Plane partition} cube stack representations adopted in the literature by Andrews\index{Andrews, G.E.} and others over the contemporary mathematical landscapes. As a mathematician already versed in the ideas of our earlier chapters, it becomes natural to ask how say, Bijection equivalences, $q$-series partition generating functions, and such, may become the same world as the statistical mechanics models involving assigned values of \textit{spin charges} along a vertex or edge of a lattice of a certain structure, shape and form. We put such questions aside for now, as we want to present Baxter's extraordinary findings here. However, these questions of equivalences of theories and analyses are worthy of deeper understanding it seems.

In seeking to calculate the free energy in the hard hexagon model\index{Hard hexagon model} as depicted in figure \ref{Fig6}, Baxter was led to consider a range of logical options or \textit{Regimes}, which are here depicted as Baxter described in 1980.

\begin{figure} [ht]
\centering
    \includegraphics[width=14.5cm,angle=0,height=8.34cm]{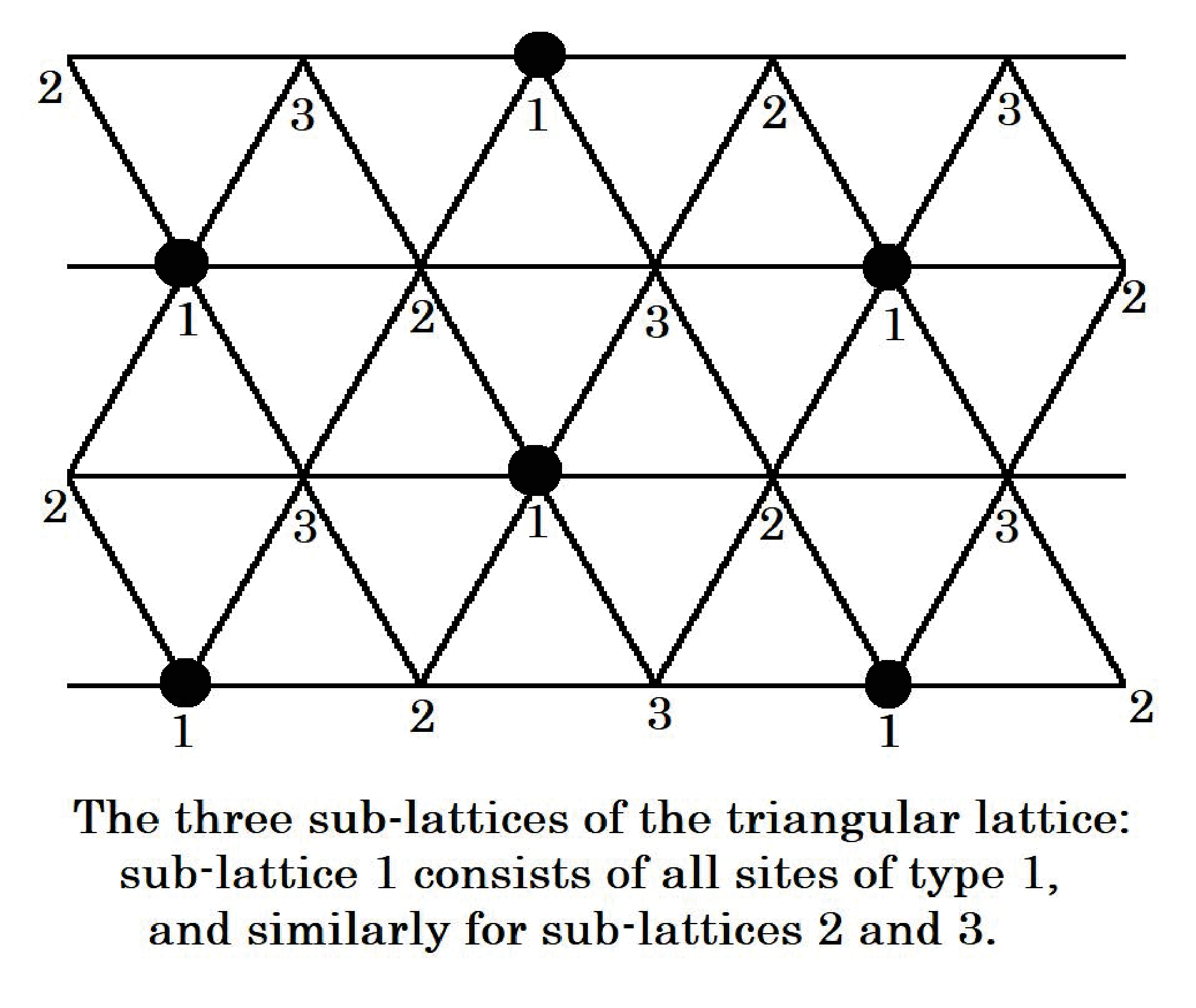}
  \caption[Baxter's three sub-lattices of the triangular lattice.]{Baxter's three sub-lattices of the triangular lattice.} \label{Fig8}
\end{figure}

The three sub-lattices of the triangular lattice: sub-lattice 1 consists of all sites of type 1, and similarly for sub-lattices 2 and 3. Adjacent sites lie on different sub-lattices. a close-packed arrangement of particles (black circles) is shown: all sites of one sub-lattice (in this case sub-lattice 1) are occupied, the rest are empty. (See figure \ref{Fig8})

\subsection{What Baxter found in 1980.}

So, continuing our discussion of equation (\ref{7.01}), we want
to calculate $Z$, or rather the partition-function per site of the infinite lattice

\begin{equation} \label{7.01a}
  \kappa = \lim_{N \rightarrow \infty} Z^{\frac{1}{N}}
\end{equation}
as a function of the positive real variable $z$. This $z$ is known as the 'activity'.
This problem can be put into what is called  `spin'-type language by associating with
each site $i$ a variable $\sigma_i$. However, instead of the usual approach in statistical mechanics models of letting $\sigma_i$ take values $1$ and $-1$, we take the values $0$ and $1$: if the site (lattice point) is empty, then $\sigma_i = 0$; if it is
full then $\sigma_i = 1$. Thus $\sigma_i$ is the number of particles at site $i$: the `occupation number'. Then (\ref{7.01}) can be written as

\begin{equation} \label{7.01b}
  Z = \sum_{\sigma} z^{(\sigma_1 + \sigma_2 + \ldots + \sigma_N)} \prod_{i,j}(1 - \sigma_i \sigma_j),
\end{equation}
where the product is over all edges $(i,j)$ of the triangular lattice, and the
sum is over all values ($0$ and $1$) of all the occupation numbers
$\sigma_1, \sigma_2, ... ,\sigma_N$.

We expect this model to undergo a phase transition from an homogeneous
fluid state at low activity $z$ to an inhomogeneous solid state at high
activity $z$.
To see this, divide the lattice into three sub-lattices $1,2,3,$ so that no
two sites of the same type are adjacent. Then there are three possible close-packed configurations of particles on the lattice: either
all sites of type 1 are occupied, or all sites of type 2, or all sites of type 3.
Suppose we fix the boundary sites as in the first possibility, i.e. all
boundary sites of type 1 are full, and all other boundary sites are empty.
Then for an infinite lattice the second and third possibilities give a negligible
contribution to the sum-over-states in (\ref{7.01b}).
Clearly, sites on different sub-lattices are no longer equivalent. Let $\rho_r$
be the local density at a site of type $r$, given by,

\begin{equation} \label{7.01c}
  \rho_r = \langle \sigma_l \rangle = Z^{-1} \sum_{\sigma} \sigma_l z^{(\sigma_1 + \sigma_2 + \ldots + \sigma_N)} \prod_{i,j}(1 - \sigma_i \sigma_j),
\end{equation}
where $l$ is a site of type $r$.

When $z$ is infinite, the system is close-packed with all sites of type $1$
occupied, so $\rho_1, = 1$, $\rho_2 = \rho_3 = 0$. We can expand each $\rho_r$ in inverse powers
of $z$ by considering successive perturbations of the close-packed state. For
a site $l$ deep inside a large lattice, this gives
\begin{eqnarray}
 \label{7.01d}  \rho_1 &=& 1 -z^{-1} -5z^{-2} -34z^{-3} -267z^{-4} - 2037z^{-5} - \ldots \\
 \nonumber      \rho_2 &=& \rho_3 = z^{-2} + 9z^{-3} + 80z^{-4} + 965z^{-5} + \ldots
\end{eqnarray}
The system is therefore not homogeneous, since $\rho_1, \rho_2, \rho_3$, are not all
equal. This contrasts with the low-activity situation: starting from the state
with all sites empty and successively introducing particles, we obtain

\begin{equation} \label{7.01e}
  \rho_1 = \rho_2 = \rho_3 = z^{1} -7z^{2} -58z^{3} -519z^{4} + 4856z^{5} - \ldots
\end{equation}
To all orders in this expansion it is true that $\rho_1 = \rho_2 = \rho_3$.

The system is therefore inhomogeneous for sufficiently large $z$, and
homogeneous for sufficiently small $z$ if the series converge. There must be a critical value $z_c$, of $z$ above which the
system ceases to be homogeneous. Since the homogeneous phase is typical
of a fluid, and the ordered inhomogeneous phase is typical of a solid, the
model can be said to undergo a fluid to solid transition at $z = z_c$.

 \begin{figure} [ht]
\centering
    \includegraphics[width=14.5cm,angle=0,height=8.34cm]{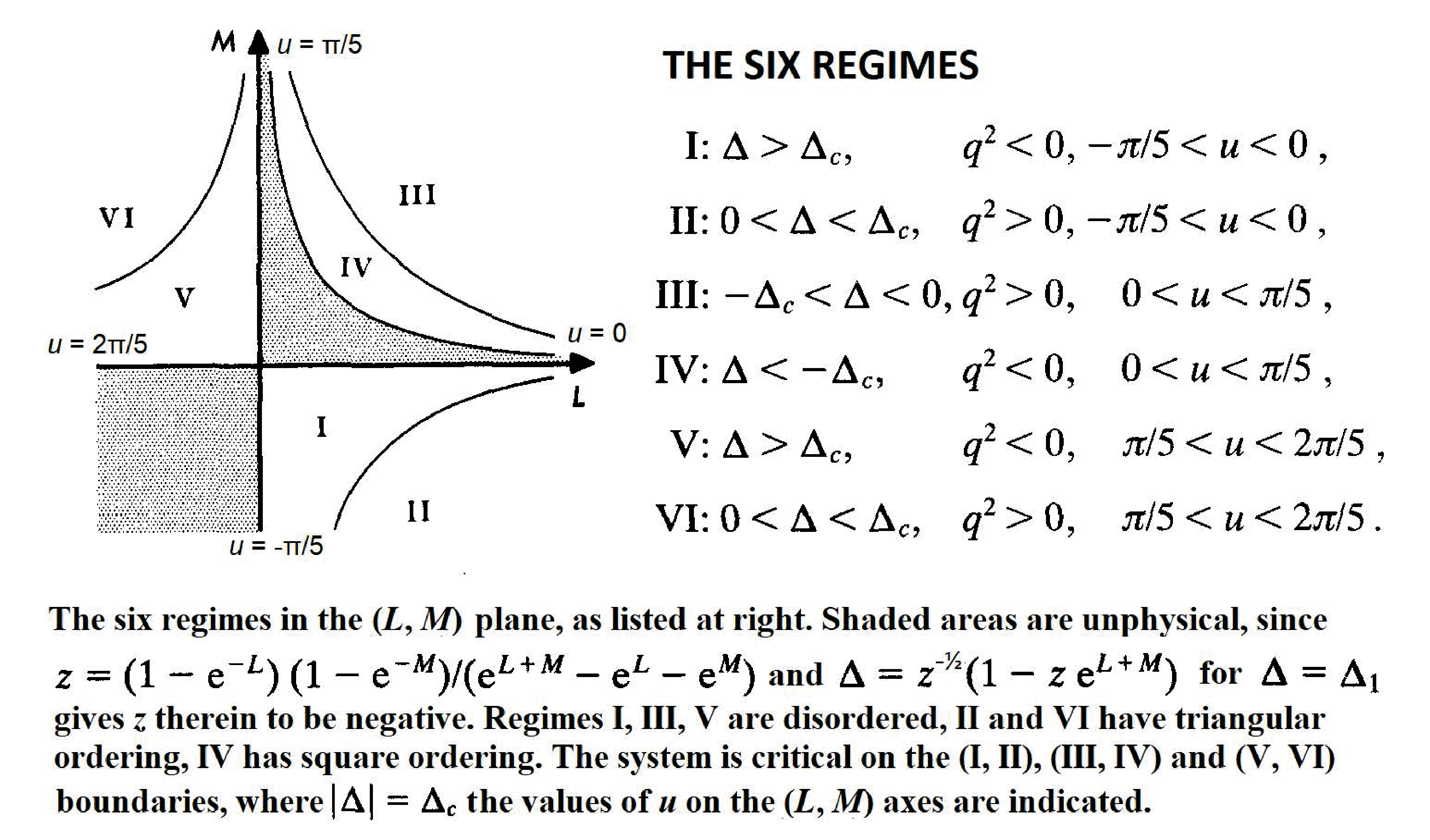}
  \caption[Baxter's six Regimes for the Hard Hexagon Model.]{Baxter's six Regimes for the Hard Hexagon Model. The six regimes in the $(L,M)$ plane. Shaded areas are not physical. Regimes I, III and V are disordered, and regimes II and VI have triangular ordering, while Regime IV has square ordering.} \label{Fig7}
\end{figure}\index{Hard hexagon model}

\subsection{Baxter's 1980 numerical preliminaries.}

Baxter then applied the Corner Transfer Matrix approach to the matrix eigenvalues $a_1, \ldots, a_{10}$, for the hard hexagon
model with $z = 1$. The values were approximate, being calculated from finite
truncations of the triangular lattice analogue of the matrix equation related to this.
The eigenvalues occur in groups of comparable magnitude, and it is sensible to
include all members of a group. For this reason the truncations used were $2 \times 2$,
$3 \times 3$, $5 \times 5$, $7 \times 7$ and $10 \times 10$. Each $a_i$ was given for successively larger truncations,
and clearly each tended rapidly to a limit. This limit is its exact value for the
infinite-dimensional corner transfer matrix. Refer to Baxter \cite[Chapter 13]{rB1982} for how the Corner Transfer Matrix context applies.

Baxter was able to determine the required coefficients up to 30 terms of the power series \textit{exact solution}. Equipped with knowledge from his famous eight-vertex model exact solution from a few years earlier, Baxter was able to put the hard hexagon proposed solution expansion into the form
\begin{equation} \label{7.01f}
  z = -x \prod_{n=1}^{\infty} (1-x^n)^{c_n}.
\end{equation}

\bigskip

With this in mind he discovered that
\begin{eqnarray}
 \nonumber   c_1, c_2, \ldots, c_{29} &=& 5, -5, -5, 5, 0, 5, -5, -5, 5, 0, \\
 \nonumber                            & & 5, -5, -5, 5, 0, 5, -5, -5, 5, 0, \\
 \label{7.01g}                        & & 5, -5, -5, 5, 0, 5, -5, -5, 5.
\end{eqnarray}
Baxter then was able to infer that
\begin{equation} \label{7.01h}
  z = -x \left( \frac{H(x)}{G(x)} \right)^5,
\end{equation}
where
\begin{eqnarray}
 \label{7.01i}  G(x) &=& \prod_{n=1}^{\infty} \frac{1}{(1-x^{5n-4})(1-x^{5n-1})} , \\
 \label{7.01j}  H(x) &=& \prod_{n=1}^{\infty} \frac{1}{(1-x^{5n-3})(1-x^{5n-2})} .
\end{eqnarray}

The mathematical partition theorist will will recognise the products $G(x)$ and $H(x)$ as from the Rogers-Ramanujan\index{Ramanujan, S.} identities\index{Rogers-Ramanujan identities}.

So, Baxter followed through this Corner Transfer Matrix approach to calculate the coefficients and determine a range of identities for each of the Hard Hexagon Regime types of allowable spin configurations. What he derived, is what Andrews\index{Andrews, G.E.} subsequently was able to prove on his visit to Baxter in Australia the following year. So, we have set the scene for Andrews\index{Andrews, G.E.} proofs in 1981.

\subsection{What Andrews found in 1981.}\index{Andrews, G.E.}

So, after Baxter's solution to the Hard Hexagon Model\index{Hard hexagon model} in 1980, the opening to a new world of research in both theoretical physics and partition-theoretic mathematics was about to happen. George E. Andrews\index{Andrews, G.E.} visited Baxter in 1981, and proved the identities arisen from the Regimes of Baxter's Hard Hexagon exact solutions.

In a landmark paper, Andrews \cite{gA1981}\index{Andrews, G.E.} goes on then to say that Baxter found the following identities occurring for the six regimes, conspicuously missing (but later returning to) the more complicated Regime II, as follows:

\textbf{Regime I}

\begin{equation} \label{7.02}
  \sum_{n=0}^{\infty} \frac{q^{n^2}}{(q)_n} = \frac{1}{\Pi(q,q^4;q^5)};
\end{equation}

\begin{equation} \label{7.03}
  \sum_{n=0}^{\infty} \frac{q^{n(n+1)}}{(q)_n} = \frac{1}{\Pi(q^2,q^3;q^5)};
\end{equation}

\bigskip

\textbf{Regime III}

\begin{equation} \label{7.04}
  \sum_{n=0}^{\infty} \frac{q^{\frac{n(3n-1)}{2}}}{(q)_n (q;q^2)_n} = \frac{\Pi(q^4,q^6,q^{10};q^{10})}{(q)_{\infty}};
\end{equation}

\begin{equation} \label{7.05}
  \sum_{n=0}^{\infty} \frac{q^{\frac{3n(n+1)}{2}}}{(q)_n (q;q^2)_{n+1}} = \frac{\Pi(q^2,q^8,q^{10};q^{10})}{(q)_{\infty}};
\end{equation}

\bigskip

\textbf{Regime IV}

\begin{equation} \label{7.06}
  \sum_{n=0}^{\infty} \frac{q^{n(n+1)}}{(q)_{2n+1}} = \frac{\Pi(q^3,q^7,q^{10};q^{10})\Pi(q^4,q^{16};q^{20})}{(q)_{\infty}};
\end{equation}

\begin{equation} \label{7.07}
  \sum_{n=0}^{\infty} \frac{q^{n(n+1)}}{(q)_{2n}} = \frac{\Pi(q,q^9,q^{10};q^{10})\Pi(q^8,q^{12};q^{20})}{(q)_{\infty}};
\end{equation}

\begin{equation} \label{7.08}
  \sum_{n=0}^{\infty} \frac{q^{n^2}}{(q)_{2n}} = \frac{1}{\Pi(q^4,q^{16};q^{20}){(q;q^2)_{\infty}}};
\end{equation}

\begin{equation} \label{7.09}
  \sum_{n=1}^{\infty} \frac{q^{n^2}}{(q)_{2n-1}} = \frac{q}{\Pi(q^8,q^{12};q^{20}){(q;q^2)_{\infty}}}.
\end{equation}

These use the standard notation of Slater \cite{lS1966},

\begin{equation} \label{7.10}
  (a)_{n} = (a;q)_{n} = \prod_{j=0}^{n-1} (1-aq^j);
\end{equation}

\begin{equation} \label{7.11}
  (a)_{\infty} = (a;q)_{\infty} = \prod_{j=0}^{\infty} (1-aq^j);
\end{equation}

\begin{equation} \label{7.12}
  \Pi(a_1,a_2,...,a_r;q) = (a_1)_{\infty} (a_2)_{\infty} ... (a_r)_{\infty}.
\end{equation}

These results either are given explicitly by Rogers (\cite{lR1894}, \cite{lR1917}) or are
immediate consequences of his work: (\ref{7.02}) is equation 1 on
page 328 of \cite{lR1894}; (\ref{7.03}) is equation 2 on page 329 of \cite{lR1894}.
(\ref{7.02}) and (\ref{7.03}) are of course the now famous Rogers-Ramanujan identities\index{Rogers-Ramanujan identities}\index{Ramanujan, S.}.

(\ref{7.04}) is equation 2, line 3, on page 330 of \cite{lR1917}; (\ref{7.05}) is equation
2, line 2, on page 330 of \cite{lR1917}; (\ref{7.06}) is implicit in the identity
of equations 2 and 3 on page 330 of \cite{lR1894} when $\lambda = 1$ and $q$ is
replaced by $-q$ (explicitly given by Slater \cite{lS1966}, equation 94);
(\ref{7.07}) is equation 13 on page 332 of \cite{lR1894}; (\ref{7.08}) is the second
equation of page 331 of \cite{lR1894}; (\ref{7.09}) is equation 3, line 2, on page
330 of \cite{lR1917}.

For \textbf{Regime II}, however, it turns out that one must consider
the following rather complicated one-dimensional partition function:

\begin{equation}\label{7.13}
 F_k(\sigma_1) =  \lim_{m \rightarrow \infty} \sum_{\sigma_2,\sigma_3,...,\sigma_{m}}
             q^{i(\sigma_{i} \sigma_{i+2} - \sigma_{i+1} + \overline{\sigma}_{i+1})}.
\end{equation}

in which the summation runs over all possible $(m - 1)$-tuples
$(\sigma_2,\sigma_3,...,\sigma_{m})$ subject to the fairly complicated and stringent conditions:

\begin{equation} \nonumber
\left\{
  \begin{array}{ll}
    \sigma_i = 0 \, or \, 1, & \hbox{$1 \leq j \leq m$;} \\
    \sigma_i + \sigma_{i+1} \leq 1, & \hbox{$1 \leq j \leq m-1$;} \\
    \sigma_m = \sigma_{m+1}=0, & \hbox{1;} \\
    \overline{\sigma}_i=1, & \hbox{$i \equiv k (mod \, 3)$;} \\
    \overline{\sigma}_i=0, & \hbox{$i \neq k (mod \, 3)$.}
  \end{array}
\right.
\end{equation}

Baxter obtains recurrence relations for refinements of these functions $F_k(\sigma_1)$;
however, the techniques that he applies successfully to solve
the recurrence relations in the other three regimes fail here.
For this reason he is unable to find counterparts of the infinite
series in (\ref{7.02}) to (\ref{7.09}). By direct expansion he obtains overwhelming
evidence to conjecture that each of $F_1(0), F_1(1), F_2(0), F_2(1), F_3(0)$, and $F_3(1)$
are identical with elegant combinations of infinite products.

Next we shall give double series expansions for the $F_k(\sigma_1)$ that establish all six of Baxter's conjectures. Apart
from their contribution to Baxter's solution of the hard-hexagon model, these results are also surprising mathematically. They
are not apparently limiting cases of known basic hypergeometric series identities; this is in contradistinction to the fact that
the place of equations (\ref{7.02}) to (\ref{7.09}) in the hierarchy of basic hypergeometric series was back then well known (see Slater \cite{lS1966}, Bailey \cite{wB1947} and \cite{wB1949}).
Here later, we shall describe the results and techniques required to establish these theorems.

\bigskip

\section{Hard Hexagon Model Regime II identities from Baxter}

So, the 1980\index{Hard hexagon model} conjectures by Baxter for Regime II were rigorously proven by Andrews\index{Andrews, G.E.} in 1981 (see \cite{gA1981}), although, as Andrews\index{Andrews, G.E.} says, the results could have been known to, or derived by, Rogers \cite{lR1917} or Schur \cite{iS1973}) in 1917, or by Slater \cite{lS1966} originally in 1952.
However, these results were not explicitly given back then, and Baxter's derivation was after all, innovative and a significant step forward in the theory. So we state the relevant identities in the following six equations.

\bigskip

\textbf{Regime II}

\begin{equation}\label{7.14}
 F_1(0) = \sum_{n=0}^{\infty} \, \sum_{0 \leq r \leq \frac{3n+1}{2}} \frac{q^{\frac{3n(n+1)}{2}-r}}{(q^2;q^2)_r (q)_{3n-2r+1}} = \frac{\Pi(q^4,q^{11},q^{15};q^{15})+ q \Pi(q,q^{14},q^{15};q^{15})}{(q)_{\infty}};
\end{equation}

\begin{equation}\label{7.15}
  F_1(1) = \sum_{n=0}^{\infty} \, \sum_{0 \leq r \leq \frac{3n}{2}} \frac{q^{\frac{3n(n+1)}{2}-r}}{(q^2;q^2)_r (q)_{3n-2r}} = \frac{\Pi(q^7,q^8,q^{15};q^{15})+ q \Pi(q^2,q^{13},q^{15};q^{15})}{(q)_{\infty}};
\end{equation}

\begin{equation}\label{7.16}
  F_2(0) = \sum_{n=1}^{\infty} \, \sum_{0 \leq r \leq \frac{3n-1}{2}} \frac{q^{\frac{n(3n-1)}{2}-r}}{(q^2;q^2)_r (q)_{3n-2r-1}} = \frac{\Pi(q^6,q^9,q^{15};q^{15})}{(q)_{\infty}}
\end{equation}
\begin{equation} \nonumber
 = 1 + q + 2 q^2 + 3 q^3 + 5 q^4 + 7 q^5 + 10 q^6 + 14 q^7 + 20 q^8 + 26 q^9 + 36 q^{10} + 47 q^{11} + 63 q^{12}
 \end{equation}
\begin{equation} \nonumber
 + 81 q^{13} + 106 q^{14} + 135 q^{15} + 174 q^{16} + 219 q^{17} + 278 q^{18} + 347 q^{19} + 436 q^{20} + O(q^{21});
\end{equation}

\begin{equation}\label{7.17}
  F_2(1) = \sum_{n=0}^{\infty} \, \sum_{0 \leq r \leq \frac{3n+1}{2}} \frac{q^{\frac{n(3n+5)}{2}+1-r}}{(q^2;q^2)_r (q)_{3n-2r+1}} =
  \frac{q \Pi(q^3,q^{12},q^{15};q^{15})}{(q)_{\infty}}
\end{equation}
\begin{equation} \nonumber
 = q + q^2 + 2 q^3 + 2 q^4 + 4 q^5 + 5 q^6 + 8 q^7 + 10 q^8 + 15 q^9 + 19 q^{10} + 27 q^{11} + 34 q^{12} + 46 q^{13}
 \end{equation}
\begin{equation} \nonumber
 + 58 q^{14} + 77 q^{15} + 96 q^{16} + 125 q^{17} + 155 q^{18} + 198 q^{19} + 244 q^{20} +  O(q^{21});
\end{equation}

\begin{equation}\label{7.18}
  F_3(0) = F_2(0) = \sum_{n=0}^{\infty} \, \sum_{0 \leq r \leq \frac{3n}{2}} \frac{q^{\frac{n(3n+1)}{2}-r}}{(q^2;q^2)_r (q)_{3n-2r}} =
  \frac{\Pi(q^6,q^9,q^{15};q^{15})}{(q)_{\infty}};
\end{equation}

\begin{equation}\label{7.19}
  F_3(1) =  F_2(1) = \sum_{n=1}^{\infty} \, \sum_{0 \leq r \leq \frac{3n-1}{2}} \frac{q^{\frac{n(3n+1)}{2}-r}}{(q^2;q^2)_r (q)_{3n-2r-1}} =
  \frac{q \Pi(q^3,q^{12},q^{15};q^{15})}{(q)_{\infty}}.
\end{equation}

So, Baxter had conjectured the identity of each of the $F_k(\sigma_i)$ with
the corresponding infinite products given above. His approach involved firstly, methods for the treatment of the expressions given in (\ref{7.12}) so that the double series representations given above can be found. Secondly, a set of transformations was required to allow
identification of the double series with the appropriate infinite product expression.

\bigskip

\section{Outline of proofs of Baxter's Regime II Conjectures}

Our methodology differs from that of Baxter immediately. Baxter's
development of series-product identities relied on the taking of
the limit as $m$ tends to $\infty$ in (\ref{7.12}). We instead find representations
for the partition functions arising in \textbf{Regime III} with
$m$ remaining fixed and finite. We then make use of the significant fact
that when $m$ is finite we can map from \textbf{Regime III}
to \textbf{Regime II} by the transformation $q \mapsto 1/q$. Our solution of
\textbf{Regime III} (on the way proving (\ref{7.01}) to (\ref{7.06}) relies
on the two following polynomial identities:

\begin{equation} \label{7.20}
  \sum_{n,r\geq 0} q^{\frac{n(3n+1}{2}} \left[ \begin{array}{c}
                                                 N-2n-2r \\
                                                 n
                                               \end{array}
   \right]_q \left[ \begin{array}{c}
                                                 r+n \\
                                                 r
                                               \end{array}
   \right]_{q^2} q^r
 =  \sum_{\lambda=-\infty}^{\infty} (-1)^\lambda q^{\lambda(5\lambda+1)} \left[ \begin{array}{c}
                                                 N \\
                                                 \left[ \frac{n-5\lambda}{2}\right]
                                               \end{array}
   \right]_q ;
\end{equation}

\begin{equation} \label{7.21}
  \sum_{n,r\geq 0} q^{\frac{n(3n+1}{2}} \left[ \begin{array}{c}
                                                 N-2n-2r-1 \\
                                                 n
                                               \end{array}
   \right]_q \left[ \begin{array}{c}
                                                 r+n \\
                                                 r
                                               \end{array}
   \right]_{q^2} q^r
 =  \sum_{\lambda=-\infty}^{\infty} (-1)^\lambda q^{\lambda(5\lambda-3)} \left[ \begin{array}{c}
                                                 N \\
                                                 \left[ \frac{n-5\lambda}{2}\right]+1
                                               \end{array}
   \right]_q ;
\end{equation}

in which

\begin{equation} \label{7.22}
  \left[ \begin{array}{c}
       N \\
       M
         \end{array}
   \right]_q
   =
   \left\{
     \begin{array}{ll}
       \frac{(1-q^N)(1-q^{N-1})\cdots (1-q^{N-M+1})}{(1-q^M)(1-q^{M-1})\cdots (1-q)}, & \hbox{for $M \geq 0$;} \\
       0, & \hbox{for $M<0, N \geq 0$,}
     \end{array}
   \right.
\end{equation}

and $[x]$ = the largest integer not exceeding $x$. If in (\ref{7.20})
we let $N \rightarrow \infty$, the first result required for \textbf{Regime III}, that being
(\ref{7.22}), is obtained. Similarly, if $N \rightarrow \infty$ in (\ref{7.23}), we obtain (\ref{7.04}).

To obtain (\ref{7.14}) to (\ref{7.19}), we replace $N$ by $3N + a$
($a = 0, \pm 1$) in (\ref{7.20}) and (\ref{7.21}), and then replace $q$ by $1/q$. Next
multiply by the minimal power of $q$ necessary to produce polynomials
in $q$, and then let $N \rightarrow \infty$. This process produces the
identities of series and products described in these (\ref{7.18}) to (\ref{7.19}), and
the relationship between \textbf{Regimes II} and \textbf{III} that follows from
the replacement of $q$ by $1/q$ resulting in the identity with the various $F_k(\sigma_l)$.

\bigskip

\section{Transforms between Baxter's Regime II and Regime III}

The results described here belie the fact that the Rogers-Ramanujan\index{Ramanujan, S.} type identities\index{Rogers-Ramanujan identities} for
\textbf{regime II} of the hard-hexagon model are now rigorously established.
On the other hand, there are numerous interesting long range
questions more of interest in the theory of partitions and
\textit{q}-series that have been extensively explored under three approaches:

\textit{Approach (i)}: Suppose the Rogers-Ramanujan\index{Ramanujan, S.} partition ideal [see Andrews\index{Andrews, G.E.} \cite[chapter 8]{gA1977} for detailed discussion of partition ideals] is replaced by another classical partition ideal; what happens in \textbf{Regimes II}, \textbf{III}, and \textbf{IV} appropriately modified?

\textit{Approach (ii)}: The $q \rightarrow 1/q$ duality of \textbf{Regimes II} and \textbf{III} also exists between
\textbf{Regimes I} and \textbf{IV}. In fact, the relevant polynomial identities for this latter relationship are

\begin{equation} \label{7.23}
  \sum_{j \geq 0} q^{j^2} \left[ \begin{array}{c}
                                                 N-j \\
                                                 j
                                               \end{array}\right]_q
 =  \sum_{\lambda=-\infty}^{\infty} (-1)^\lambda q^{\frac{\lambda(5\lambda+1)}{2}}
                                               \left[ \begin{array}{c}
                                                 N \\
                                                 \left[ \frac{N-5\lambda}{2}\right]
                                               \end{array}
                                               \right]_q ;
\end{equation}

\begin{equation} \label{7.24}
  \sum_{j \geq 0} q^{j(j+1)} \left[ \begin{array}{c}
                                                 N-j \\
                                                 j
                                               \end{array}\right]_q
 =  \sum_{\lambda=-\infty}^{\infty} (-1)^\lambda q^{\frac{\lambda(5\lambda-3)}{2}}
                                               \left[ \begin{array}{c}
                                                 N \\
                                                 \left[ \frac{N+1-5\lambda}{2}\right] +1
                                               \end{array}
                                               \right]_q .
\end{equation}

These identities were completely stated in Andrews \cite{gA1970}\index{Andrews, G.E.} and have their origin in the work of Schur \cite{iS1973a}. The arguments used to obtain (\ref{7.14}) to (\ref{7.19}) from (\ref{7.20}) and (\ref{7.21}) may now be turned on (\ref{7.23}) and (\ref{7.24}) to obtain (\ref{7.06}) to (\ref{7.09}), a relationship previously unnoticed. Thus a \textit{duality theory} between various sets of identities of the Rogers-Ramanujan\index{Rogers-Ramanujan identities}\index{Ramanujan, S.} type presents for exploration.

\textit{Approach (iii)}: The analytic duality described above has a corresponding manifestation in the partition-theoretic interpretations of the various identities considered. Thus the well-known combinatorial interpretations of (\ref{7.02}) and (\ref{7.03}) are \textit{dual} to the combinatorial interpretations of (\ref{7.06}) to (\ref{7.09}). Since the 1980s this duality has been developed in two distinct streams, via statistical mechanics Integrable Systems\index{Integrable function} and in the theory of mathematical integer partitions.

\bigskip

\section{The partition function of the Hard Hexagon Model}

The hard hexagon model\index{Hard hexagon model} occurs within the framework of the grand canonical ensemble, where the total number of particles (the \textit{hexagons}) is allowed to vary naturally, and is fixed by a chemical potential. In the hard hexagon model, all valid states have zero energy, and so the only important thermodynamic control variable is the ratio of chemical potential to temperature $\frac{\mu}{\kappa T}$. The exponential of this ratio,
 $z = \exp \left(\frac{\mu}{\kappa T}\right)$ is called the activity and larger values correspond roughly to denser configurations.

For a triangular lattice with $N$ sites, the grand partition function is

\begin{equation}\label{7.25}
  Z(z) = \sum_{n} z^n g(n,N) - 1 + Nz + \frac{1}{2}N(N-7)z^2 + \cdots
\end{equation}

where $g(n, N)$ is the number of ways of placing $n$ particles on distinct lattice sites such that no two are adjacent. The function $\kappa$ is defined by

\begin{equation}\label{7.26}
  \kappa(z) = \lim {Z(z)}^{1/N} = 1 + z - 3 z^2 + \cdots
\end{equation}

so that $\log(\kappa)$ is the free energy per unit site. Solving the hard hexagon model\index{Hard hexagon model} means (roughly) finding an exact expression for $\kappa$ as a function of $z$.

The \textbf{mean density} $\rho$ is given for small $z$ by

\begin{equation} \label{7.27}
  \rho = z \frac{d \log(\kappa)}{dz} = z - 7 z^2 + 58 z^3 - 519 z^4 + 4856 z^5 - \cdots .
\end{equation}

The vertices of the lattice fall into three classes numbered 1, 2, and 3, given by the 3 different ways to fill space with hard hexagons. There are 3 local densities $\rho_1, \rho_2, \rho_3$, corresponding to the 3 classes of sites. When the activity is large the system approximates one of these 3 packings, so the local densities differ, but when the activity is below a critical point the three local densities are the same. The critical point separating the low-activity homogeneous phase from the high-activity ordered phase is

\begin{equation}\nonumber
z_{c}=\frac{(11+5{\sqrt {5}})}{2} = \phi^5 = 11.09017\ldots
\end{equation}

with golden ratio $\phi = \frac{1 + \sqrt(5)}{2}$. Above the critical point the local densities differ and in the phase where most hexagons are on sites of type 1 can be expanded as

\begin{equation}\nonumber
  \rho_1 = 1 - z^{-1} - 5 z^{-2} - 34 z^{-3} -267 z^{-4}- 2037 z^{-5} - \cdots ,
\end{equation}
\begin{equation}\nonumber
  \rho_2 = z^{-2} - 9 z^{-3} + 80 z^{-4} + 965 z^{-5} - \cdots .
\end{equation}

\subsection{Solution}\index{Hard hexagon model!Solution}

The solution is given for small values of $z < z_c$ by

\begin{equation}\nonumber
  z = \frac{-x H(x)^5}{G(x)^5},
\end{equation}
\begin{equation}\nonumber
  \kappa = \frac{H(x)^3 Q(x^5)^2}{G(x)^2} \prod_{n\geq1}\frac{(1-x^{6n-4})(1-x^{6n-3})^2(1-x^{6n-2})}{(1-x^{6n-5})(1-x^{6n-1})(1-x^{6n})^2},
\end{equation}
\begin{equation}\nonumber
  \rho = \rho_1 = \rho_2 = \rho_3 = \frac{-x G(x) H(x^6) P(x^3)}{P(x)},
\end{equation}

where

\begin{equation}\nonumber
  G(x) = \prod_{n\geq1} \frac{1}{(1-x^{5n-4})(1- x^{5n-1})},
\end{equation}
\begin{equation}\nonumber
  H(x) = \prod_{n\geq1} \frac{1}{(1-x^{5n-3})(1- x^{5n-2})},
\end{equation}
\begin{equation}\nonumber
  P(x) = \prod_{n\geq1} (1-x^{2n-1}) = \frac{Q(x)}{Q(x)^2},
\end{equation}
\begin{equation}\nonumber
  Q(x) = \prod_{n\geq1} (1-x^{n}).
\end{equation}

For large $z > z_c$ the solution (in the phase where most occupied sites have type 1) is given by

\begin{equation}\nonumber
  z = \frac{G(x)^5}{x H(x)^5},
\end{equation}
\begin{equation}\nonumber
  \kappa = x^{\frac{1}{3}} \frac{G(x)^3 Q(x^5)^2}{H(x)^2} \prod_{n\geq1}\frac{(1-x^{3n-2})(1-x^{3n-1})}{(1-x^{3n})^2},
\end{equation}
\begin{equation}\nonumber
  \rho_1 = \frac{H(x)Q(x)(G(x)Q(x)+x^2H(x^9)Q(x^9))}{Q(x^3)^2},
\end{equation}
\begin{equation}\nonumber
  \rho_2 = \rho_3 = \frac{x^2 H(x)Q(x)H(x^9)Q(x^9)}{Q(x^3)^2},
\end{equation}
\begin{equation}\nonumber
  R = \rho_1 - \rho_2 = \frac{Q(x)Q(x^5)}{Q(x^3)^2}.
\end{equation}

The functions $G$ and $H$ turn up in the Rogers-Ramanujan\index{Ramanujan, S.} identities, and the function $Q$ is the Euler function\index{Euler, L.}, which is closely related to the Dedekind eta function. If $x = e^{2 \pi i \tau}$, then $q^{-1/60}G(x)$, $x^{11/60}H(x)$, $x^{-1/24}P(x)$, $z, \kappa, \rho, \rho_1, \rho_2$, and $\rho_3$ are modular functions of $\tau$, while $x^{1/24}Q(x)$ is a modular form of weight $1/2$. Since any two modular functions are related by an algebraic relation, this implies that the functions $\kappa, z, R, \rho$ are all algebraic functions of each other of quite high degree.

\section{Rogers-Ramanujan shift from Mathematics to Physics}\index{Ramanujan, S.}

In 1998 the physicists Alexander Berkovich and Barry M. McCoy\index{McCoy, B.M.}\index{Berkovich, A.} from State University of New York wrote an expository paper \cite{aB1998} explaining how the story of the Rogers-Ramanujan identities\index{Rogers-Ramanujan identities}\index{Ramanujan, S.} transitioned from mathematics to physics. Put simply, in 1894 L.J. Rogers \cite{lR1894} proved the following identities for $a = 0, 1$ between infinite series and products valid for $|q| < 1$,
\begin{equation}  \label{7.28}
  \sum_{n=0}^{\infty} \frac{q^{n(n+a)}}{(q)_n}=\prod_{n=1}^{\infty} \frac{1}{(1-q^{5n-1-a})(1-q^{5n-4+a})}
\end{equation}
\begin{equation*}
  = \frac{1}{(q)_\infty} \sum_{m=-\infty}^{\infty} (q^{n(10n+1+2a)} - q^{(5n+2-a)(2n+1)}),
\end{equation*}
where $(q)_n=\prod_{j=0}^{n} (1-q^j)$.

A fascinating quote from Berkovich and McCoy\index{McCoy, B.M.}\index{Berkovich, A.} on Rogers-Ramanujan identities\index{Rogers-Ramanujan identities}\index{Ramanujan, S.}:

\bigskip

\textit{Think About It...}

\begin{quote}
“For about the first 85 years after their discovery interest in these identities and their generalizations was confined to mathematicians. Many ingenious proofs and relations applying combinatorics, basic hypergeometric functions and Lie algebras were discovered by MacMahon\index{MacMahon, P.A.}, Rogers, Schur, Ramanujan\index{Ramanujan, S.}, Watson, Bailey, Slater, Gordon, G\"{o}llnitz, Andrews,\index{Andrews, G.E.} Bressoud\index{Bressoud, D.M.}, Lepowsky and Wilson; so by 1980 there were over 130 isolated identities and several infinite families of identities known.

\bigskip

The entry of these identities into physics occurred in the early 1980s when Baxter \cite{rB1981}, Andrews,\index{Andrews, G.E.} Baxter and Forrester (see \cite{gA1981} and \cite{pF1985}), and the Kyoto group\index{Kyoto group} \cite{eD1987} encountered (\ref{7.28}) and various generalizations in the computation of order parameters of certain lattice models of statistical mechanics.

\bigskip

A further glimpse of the relation to physics is seen in the development of conformal field theory by Belavin, Polyakov and Zamolodchikov \cite{aB1984} and the form of computation of characters of representations of Virasoro algebra by Kac \cite{vK1979}, Feigin and Fuchs \cite{bF1983} and Rocha-Caridi \cite{aR1985}. The occurrence of (\ref{7.28}) in this context led Kac \cite{vK1992} to suggest that \textit{every modular invariant representation of Virasoro should produce a Rogers-Ramanujan type identity.}''\index{Rogers-Ramanujan identities}\index{Ramanujan, S.}

\bigskip

\textit{- Alexander Berkovich and Barry M. McCoy, State University of New York, 1998}\index{Quotes:Think about it!Berkovich, A. and McCoy, B.M.}\index{Baxter, R.J.}\index{McCoy, B.M.}\index{Berkovich, A.}
\end{quote}

\bigskip

The full relation between physics and Rogers-Ramanujan identities\index{Rogers-Ramanujan identities}\index{Ramanujan, S.} is more extensive than we may think from these first indications. Starting in 1993 both Berkovich and McCoy\index{McCoy, B.M.}\index{Berkovich, A.} (see for example \cite{rK1993,rK1993a}, \cite{aB1994,aB1996,aB1998,aB1998a}, and \cite{gA1998}) have fused the physical insight of solvable lattice models in statistical mechanics with the classical work of the first 85 years and the recent developments in conformal field theory to greatly enlarge the theory of Rogers-Ramanujan identities\index{Rogers-Ramanujan identities}\index{Ramanujan, S.}. We can summarize the results of this work and present some of the current results. Our point of view for the rest of this chapter will be influenced by the statistical mechanics, but we will try to indicate where alternative viewpoints exist. Hopefully some language differences between physicists and mathematicians can be understood.

The work of the last 30 years arising in physics problems has given a new context and point of view in the study of Rogers-Ramanujan identities\index{Rogers-Ramanujan identities}\index{Ramanujan, S.}. The emphasis is not the same as in the earlier mathematical investigations and thus it is good to discuss this before giving detailed results.

In mathematics, Rogers-Ramanujan identities\index{Rogers-Ramanujan identities}\index{Ramanujan, S.} always were seen as \textit{partitions of integers equivalence theorems} or as \textit{generating function algebraic identities for integer partitions}. This latter mostly took the form of $q$-series identities where one side was a sum of $(q)_n$ terms and the other side was an infinite product of $(q)_n$ terms. In physics the discussion still centered around partitions, but now defined by terms of \textit{statistical mechanical and conformal field theory applications}. In physics nowadays in most cases where we have generalizations of the identities between the two sums, a product form is not known. So in physics, a Rogers-Ramanujan\index{Ramanujan, S.} identity means the equality of the sums without further reference to possible product forms.

What is interesting about the two diverging approaches to this topic, is that underlying the mathematics and the physics is a universality that embraces both approaches. However, for now, the two approaches not just to Rogers-Ramanujan identities\index{Rogers-Ramanujan identities}\index{Ramanujan, S.}, but to the theory of partitions, are distinct, but equally valid.

The full range of Rogers-Ramanujan identities\index{Rogers-Ramanujan identities}\index{Ramanujan, S.} is by no means yet understood and it is anticipated that both in the mathematics and in the physics there is much still left to be discovered. The physics has grown into innumerable research areas due to the now well-established fact that the Yang-Baxter equation has long been recognised as the masterkey to integrability\index{Integrable function}, providing the basis for exactly solved models which capture the fundamental physics of a number of realistic classical and quantum systems. The theory of partitions has therefore become an essential tool for physicists involved with exactly solvable models, of which there is now, including the hard-hexagon model, the Heisenberg spin chain\index{Heisenberg model}, the transverse quantum Ising chain\index{Ising model}, a spin ladder model\index{Spin ladder model}, the Lieb-Liniger Bose gas\index{Lieb-Liniger Bose gas model}, the Gaudin-Yang Fermi gas\index{Gaudin-Yang Fermi gas model} and the two-site Bose-Hubbard\index{Bose-Hubbard model} model. For an eloquent review of this in the context of \textit{condensed matter to ultracold atoms} see Batchelor and Foerster \cite{mB2016}.

\section{Partition generating functions in Physics are polynomial generalizations}

The second insight which is also present in the very first papers on the connection of Rogers-Ramanujan identities\index{Rogers-Ramanujan identities}\index{Ramanujan, S.} with physics, (see Baxter \cite{rB1981}, Andrews,\index{Andrews, G.E.} Baxter and Forrester \cite{gA1984a}, and Forrester and Baxter \cite{pF1985}) is the fact that the physics will often lead to polynomial identities, with an order depending on an integer $L$, which yield infinite series identities as $L \rightarrow \infty$.

The polynomial generalization of (\ref{7.28}) is the identity first proven by Andrews\index{Andrews, G.E.} \cite{gA1970} in 1970,

\begin{equation}  \label{7.29}
  F_a(L,q) = B_a(L,q)
\end{equation}
where
\begin{equation}  \label{7.30}
  F_a(L,q) = \sum_{n=0}^{\infty} q^{n(n+a)} \left[ \begin{array}{c}
                                                    L-n-a \\
                                                    n
                                                  \end{array}
                                            \right]_q
\end{equation}
and
\begin{equation}  \label{7.31}
  B_a(L,q) = \sum_{n=-\infty}^{\infty} (-1)^n q^{n(5n+1+2a)/2}
                                                 \left[ \begin{array}{c}
                                                    L \\
                                                    \lfloor\frac{1}{2}(L-5n-a)\rfloor
                                                  \end{array}
                                                 \right]_q
\end{equation}
where $\lfloor x \rfloor$ denotes the integer part of $x$ and the Gaussian polynomials\index{Gauss, C.F.!Gaussian polynomial} ($q$-binomial coefficients) are defined for integer $m$, $n$ by
\begin{equation}  \label{7.32}
  \left[ \begin{array}{c}
                n \\
                m
         \end{array}
  \right]_q
  = \left\{
      \begin{array}{ll}
        \frac{(q)_n}{(q)_m (q)_{n-m}}, & \hbox{$0 \leq m \leq n$;} \\
        0, & \hbox{otherwise.}
      \end{array}
    \right.
\end{equation}

The identity (\ref{7.29}) is obtained by using
\begin{equation*}
  \lim_{n \rightarrow \infty} \left[ \begin{array}{c}
                n \\
                m
         \end{array}
  \right]_q = \frac{1}{(q)_m}.
\end{equation*}
So (\ref{7.29}) is a generalization of the identity (\ref{7.28}) which we will call a Rogers-Ramanujan identity\index{Rogers-Ramanujan identities}\index{Ramanujan, S.}.

There are many known generalizations of $F_a(L, q)$ that are written in terms of the following function due to Kedem et al. \cite{rK1993a}
\begin{equation}  \label{7.33}
  f  = \sum_{restrictions} q^{\frac{1}{2} \mathbf{mBm} - \frac{1}{2} \mathbf{Am}} \prod_{\alpha=1}^{n}
    \left[ \begin{array}{c}
                ((1-\mathbf{B})\mathbf{m} + \frac{\mathbf{u}}{2})_\alpha \\
                m_\alpha
         \end{array}
    \right]_q
\end{equation}
where $\mathbf{m}$, $\mathbf{u}$ and $\mathbf{A}$ are $\mathbf{n}$ dimensional vectors and $\mathbf{B}$ is an $n \times n$ dimensional matrix and the sum is over all values of the variables $m_\alpha$ possibly subject to some restrictions (such as being even or odd). In many cases the $q$-binomials are defined by (\ref{7.32}) but there do occur cases with an extended definition
\begin{equation}  \label{7.34}
  \left[ \begin{array}{c}
                m+n \\
                m
         \end{array}
  \right]_q
  =
\left\{
  \begin{array}{ll}
    (q^{n+1})_m, & \hbox{for $m \geq 0$, $n$ integers;} \\
    0, & \hbox{otherwise.}
  \end{array}
\right.
\end{equation}
This brings (\ref{7.32}) to cover negative $n$.

The function (\ref{7.33}) is regarded as the partition function for a collection of $n$ different species of free massless (right moving) fermions with a linear energy momentum relation $e(P_j^\alpha)_\alpha = vP_j^\alpha$ where the momenta are quantized in units of $2\pi/M$ and are chosen from the sets
\begin{equation} \label{7.35}
  P_j^\alpha \in \left\{P_{\min}^\alpha,P_{\min}^\alpha + \frac{2\pi}{M},P_{\min}^\alpha + \frac{4\pi}{M},
             P_{\min}^\alpha + \frac{6\pi}{M},\ldots,P_{\max}^\alpha \right\}
\end{equation}
where
\begin{equation}  \nonumber
  P_{\min}^\alpha(\mathbf{m}) = \frac{\pi}{M} \left[ ((\mathbf{B}-1)\mathbf{m})_\alpha - \mathbf{A}_\alpha + 1  \right]
\end{equation}
and
\begin{equation}  \nonumber
  P_{\max}^\alpha(\mathbf{m}) = -P_{\min}^\alpha(\mathbf{m})+ \frac{\pi}{M} \left(\frac{\mathbf{u}}{2}-\mathbf{A}\right)_\alpha
\end{equation}
with the Fermi exclusion rule $P_{j}^\alpha \neq P_{k}^\alpha$ for $j \neq k$ and all $\alpha=1,2,\ldots,n$.

\section{The Elliptic \MakeLowercase{q}-Gamma Function}\index{Elliptic $q$-gamma function}

The elliptic gamma function\index{Elliptic $q$-gamma function} is a generalization of the $q$-gamma function\index{q-gamma function}, which is itself the $q$-analog
of the usual gamma function\index{Gamma function}. It was first examined in detail by Ruijsenaars in 1997 as an integrable function\index{Integrable function} (see \cite{sR1997}), and can be expressed in terms of the triplegamma function\index{Triplegamma function}. It is defined by
\begin{equation} \label{7.36}
  \Gamma(z;p,q) = \prod_{m=0}^{\infty} \prod_{n=0}^{\infty} \frac{1 - p^{m+1}q^{n+1}/z}{1 - p^{m}q^{n}z}.
\end{equation}
The following relation is easily seen to apply
\begin{equation} \label{7.37}
  \Gamma(z;p,q) \, \Gamma(pq/z;p,q) = 1.
\end{equation}
Using the $q$-theta function\index{q-theta function} $\theta(z;q)= \prod_{n=0}^{\infty}(1-q^nz)(1-q^{n+1}/z)$, it is well-known that
\begin{eqnarray}
\nonumber  \Gamma(pz;p,q) &=& \theta(z;q) \Gamma(z;p,q), \\
\nonumber  \Gamma(qz;p,q) &=& \theta(z;p) \Gamma(z;p,q).
\end{eqnarray}
At $p=0$ we have $\Gamma(z;0,q) \, (z,q)_\infty = 1$ using the infinite $q$-Pochhammer notation.

As with the usual gamma function, there is a duplication, a triplication and multiplication formula. We need firstly to define the function
\begin{equation} \label{7.38}
    \widetilde{\Gamma}(z;p,q) := \frac{(q,q)_\infty}{(p,p)_\infty} (\theta(q,p))^{1-z}
  \prod_{m=0}^{\infty} \prod_{n=0}^{\infty} \frac{1 - p^{m+1}q^{n+1-z}}{1 - p^{m}q^{n+z}}.
\end{equation}
We follow the approach by Felder and Varchenko \cite{gF2002} who proved the
\begin{theorem}  \label{7.39} If $r=q^n$ then
  \begin{equation} \label{7.40}
    \widetilde{\Gamma}(nz;p,q) \widetilde{\Gamma}(\frac{1}{n};p,r) \widetilde{\Gamma}(\frac{2}{n};p,r) \cdots \widetilde{\Gamma}(\frac{n-1}{n};p,r) \quad \quad \quad \quad
\end{equation}
  \begin{equation} \nonumber
\quad  \quad \quad \quad  = \left(\frac{\theta(r;p)}{\theta(q;p)}\right)^{nz-1}
     \widetilde{\Gamma}(z;p,q) \widetilde{\Gamma}(z+\frac{1}{n};p,r) \cdots \widetilde{\Gamma}(z+\frac{n-1}{n};p,r).
\end{equation}
\end{theorem}

The two most significant corollaries of theorem \ref{7.39} are as follows:

\begin{corollary}  \label{7.41} \textit{The Duplication Formula.}
  \begin{equation} \label{7.42}
    \widetilde{\Gamma}(2z;p,q) \widetilde{\Gamma}(\frac{1}{2};p,q^2)
      = \left(\frac{\theta(q^2;p)}{\theta(q;p)}\right)^{2z-1} \widetilde{\Gamma}(z;p,q^2) \widetilde{\Gamma}(z+\frac{1}{2};p,q^2).
\end{equation}
\end{corollary}
\begin{corollary}  \label{7.43} \textit{The Triplication Formula.}
  \begin{equation} \label{7.44}
    \widetilde{\Gamma}(3z;p,q) \widetilde{\Gamma}(\frac{1}{3};p,q^3)\widetilde{\Gamma}(\frac{2}{3};p,q^3)
      = \left(\frac{\theta(q^3;p)}{\theta(q;p)}\right)^{3z-1} \widetilde{\Gamma}(z;p,q^3) \widetilde{\Gamma}(z+\frac{1}{3};p,q^3)
      \widetilde{\Gamma}(z+\frac{2}{3};p,q^3).
\end{equation}
\end{corollary}

The above theorem with the Duplication and Triplication formula cases may be worthwhile examining and comparing with the functional equations\index{Functional equation} given in chapter 15 of the book by Campbell \cite{gC2024} soon to appear. For example, is there a connection applicable between the Multiplication Formula of the Elliptic Gamma Function\index{Elliptic $q$-gamma function} and the functions cited, namely,
\begin{definition}
  Define the function $F_n (t)$ for all complex numbers $a,t$
with $|a|,|t|<1$, and for $n\geq 1$ with $|q_1|,|q_2|,...,|q_n|<1$ by the sequence of functions $F_n(t)$ with
\begin{equation}\label{7.45}
  F_0(-;a,t)= \frac{1-at}{1-t}
\end{equation}
and for all of $\left|q_1\right|,\left|q_2\right|,\left|q_3\right|,...,\left|q_n\right|<1$, by
\begin{equation}\label{7.46}
  F_n(q_1,q_2,q_3,...,q_n;a,t)= \prod_{\alpha_1,\alpha_2,\alpha_3,...,\alpha_n \geq0}\frac{1-q_1^{\alpha_1} q_2^{\alpha_2} ... q_n^{\alpha_n}at}{1-q_1^{\alpha_1} q_2^{\alpha_2} ... q_n^{\alpha_n}t} \equiv \sum_{k=1}^\infty {_{n}}{A_{k}} t^k
\end{equation}
where $\alpha_1,\alpha_2,\alpha_3,...,\alpha_n$ may be all zero.
\end{definition}
We now state a functional equation\index{Functional equation} theorem for $F_n(t)$, next highlighting the 2D, 3D and 4D cases as examples in a corollary.
\begin{theorem} (see Campbell \cite[Chapter 15]{gC2024})\index{Functional equation}
  If the left side of (\ref{7.46}) is for the moment considered only as a function, $F_n(t)$, of $t$, then the following finite product holds,
  \begin{equation}\label{7.47}
    \frac{1-at}{1-t} = \frac{F_n(t)\left(\prod_{S_2}F_n(q_{i_1} q_{i_2} t)\right) \left(\prod_{S_4} F_n(q_{i_1} q_{i_2} q_{i_3} q_{i_4} t)\right) ... }
    {\left(\prod_{S_1}F_n(q_{i_1} t)\right) \left(\prod_{S_3} F_n(q_{i_1} q_{i_2} q_{i_3} t)\right) \left(\prod_{S_5}F_n(q_{i_1} q_{i_2} q_{i_3} q_{i_4} q_{i_5} t)\right) ...  }
  \end{equation}
  the numerator products each being over the $n$th order even symmetric combinations whilst the denominator products are over the $n$th order odd symmetric combinations of the variables.  We have used the notation $S_1, S_2, S_3, S_4, ...$ to denote the respective $n$th  order symmetric function product of terms.
\end{theorem}

It is worth noting that we have used the abbreviation $F(t)=F_n(q_1,q_2,q_3,...,q_n;a,t)$
and the right side of (\ref{7.47}) is a finite product of such functions going up to the $n$th order symmetric function product of terms. It is also seen here that (\ref{7.47}) is a functional equation\index{Functional equation} for a \textit{generalized right side of (\ref{7.46})} form of the $q$-binomial product. As an illustration we next give the $n=1$ up to $n=4$ examples of the functional equation\index{Functional equation} (\ref{7.46}), with $q_1 \rightarrow x$, $q_2 \rightarrow y$, $q_3 \rightarrow z$. With these variables, we assert

\begin{corollary}  (see Campbell \cite[Chapter 15]{gC2024})\index{Functional equation}
  If for $\left|w\right|<1, \left|x\right|<1, \left|y\right|<1, \left|z\right|<1$, and all complex numbers $a$ and $t$,
  \begin{eqnarray}
    \nonumber  F_1(t)=F_1(z;a,t) &=& \prod_{j\geq0} \frac{1-z^j at}{1-z^j t},  \\
    \nonumber  F_2(t)=F_2(y,z;a,t) &=& \prod_{j,k\geq0} \frac{1-y^j z^k at}{1-y^j z^k t},  \\
   \nonumber  F_3(t)=F_3(x,y,z;a,t) &=& \prod_{j,k,l\geq0} \frac{1-x^j y^k z^l at}{1-x^j y^k z^l t}, \\
    \nonumber  F_4(t)=F_4(w,x,y,z;a,t) &=& \prod_{j,k,l,m\geq0} \frac{1-w^j x^k y^l z^m at}{1-w^j x^k y^l z^m t}, \\
    etc.
    \end{eqnarray}
  then
  \begin{eqnarray}
 \nonumber  \frac{1-at}{1-t} &=& \frac{F_1(t)}{F_1(zt)},  \\
 \nonumber   \frac{1-at}{1-t} &=& \frac{F_2(t)F_2(yzt)}{F_2(yt)F_2(zt)},  \\
 \nonumber   \frac{1-at}{1-t} &=& \frac{F_3(t)F_3(yzt)F_3(xyt)F_3(xzt)}{F_3(xt)F_3(yt)F_3(zt)F_3(xyzt)}, \\
\nonumber    \frac{1-at}{1-t} &=& \frac{F_4(t)F_4(yzt)F_4(xyt)F_4(xzt)F_4(wxt)F_4(wyt)F_4(wzt)F_4(wxyzt)}
    {F_4(wt)F_4(xt)F_4(yt)F_4(zt)F_4(wxyt)F_4(wxzt)F_4(wyzt)F_4(xyzt)}, \\
    etc.
  \end{eqnarray}
\end{corollary}

The above equations are suggestive of three as yet undeveloped theories:
\begin{enumerate}
  \item[1) ] The functional equations\index{Functional equation} for generating functions imply recurrence relations for the nD vector partitions themselves. These are higher dimensional extensions of the Ramanujan\index{Ramanujan, S.} integer partition congruences.\index{Congruences of partitions}\index{Partition congruences} See Bruinier \cite{jB2002}, and Bruinier et al. \cite{jB2004a} to \cite{jB2013a} for unrestricted partition congruences.
  \item[2) ] An $n$D $m$-ary vector partition congruence theory; extending the binary partition\index{Binary partition}\index{Binary partition!Congruences} congruences of R\"{o}dseth and Gupta. See Alkauskas \cite{gA2003} for an account of binary integer congruences.\index{Congruences of partitions}\index{Partition congruences}
  \item[3) ] The $n$D generating functions suggest their own versions of Duplication Formulas, Triplication formulas, and Multiplication formulas, similar to those stated here for the Elliptic Gamma function\index{Elliptic $q$-gamma function}. This function is \textit{integrable}\index{Integrable function} and therefore related to existing theories in Statistical Mechanics, Yang-Baxter equations\index{Yang-Baxter equation}, Boltzmann weights and phase transitions for Lattice Models. (See the original 1997 account by Ruijsenaars \cite{sR1997}, and more recently the 2016 physics in Bazhanov et al. \cite{vB2016} for example.
\end{enumerate}

This narrative adjourns to future discussions on topics of functional equations\index{Functional equation}, integrable functions\index{Integrable function}, vector partition congruences\index{Vector partition!Congruences}\index{Congruences of partitions}\index{Partition congruences} and nD Lattice Models. The Elliptic Gamma Function\index{Binary partition} is a 2D example of an nD function that may \textit{play out} in a similar way in higher Euclidean spaces. The Elliptic Gamma Function seems useful for vector partition concepts not yet articulated. Also, our chapter 15 of Campbell \cite[Chapter 15]{gC2024} on Vector Partition generating functions and functional equations\index{Functional equation} may be related to solvable models associated with Yang-Baxter equations\index{Yang-Baxter equation}. This topic brings us deeply into the physics of Statistical Mechanics, but is beyond the present scope of our book. It may bring future researchers to consider cases of single or multiple spin degrees of freedom at each site of a higher dimensional lattice. The Yang-Baxter equation\index{Yang-Baxter equation} for such models may reduce to particular simple forms called Star-Triangle relations\index{Star-Triangle relation}. For a 2023 contemporary appraisal of Yang-Baxter equations related to Brauer algebras see Ca\~{n}ada \cite{aC2023}.

\bigskip

\textit{Think About It...}

\begin{quote}
The 1982 book by Rodney Baxter \textit{Exactly solved models in statistical mechanics}\index{Exactly solved model} has over 7,180 citations as at early 2023.

\bigskip

These citations are probably split approximately evenly between physics and mathematics research articles and books; the breakup as follows: Highly Influential Citations = 919; Background Citations = 2,656; Methods Citations = 1,857; Results Citations = 125.

\bigskip

\textit{- Semantic Scholar }{semanticscholar.org, February 2023.}\index{Quotes:Think about it!Baxter, R.J.}\index{Baxter, R.J.}\index{Semantic Scholar}\index{Quotes:Think about it!Semantic Scholar}
\end{quote}

\end{document}